\documentclass[aps,pre,preprint,groupedaddress]{revtex4-1}
\usepackage{graphicx}
\usepackage{epstopdf, epsfig}
\usepackage{amsmath}
\usepackage{float}
\usepackage{bigints}
\usepackage{relsize}

\begin{document}


\title{Front Roughening of Flames in Discrete Media}

\author{Fredric Lam}
\affiliation{Department of Mechanical Engineering, McGill University, Montreal, Quebec, Canada}

\author{XiaoCheng Mi}
\affiliation{Department of Mechanical Engineering, McGill University, Montreal, Quebec, Canada}

\author{Andrew J. Higgins}
\email[Corresponding author: ]{andrew.higgins@mcgill.ca}
\affiliation{Department of Mechanical Engineering, McGill University, Montreal, Quebec, Canada}


\date{\today}

\begin{abstract}
The morphology of flame fronts propagating in reactive systems comprised of randomly positioned, point-like sources is studied.  The solution of the temperature field and the initiation of new sources is implemented using the superposition of the Green's function for the diffusion equation, eliminating the need to use finite difference approximations.  The heat released from triggered sources diffuses outward from each source, activating new sources and enabling a mechanism of flame propagation.  Systems of $40000$ sources in a $200$ by $200$ two-dimensional domain were tracked using computer simulations, and statistical ensembles of $120$ realizations of each system were averaged to determine the statistical properties of the flame fronts.  The reactive system of sources is parameterized by two non-dimensional values:  The heat release time (normalized by interparticle diffusion time) and the ignition temperature (normalized by adiabatic flame temperature).  These two parameters were systematically varied for different simulations to investigate their influence on front propagation.  For sufficiently fast heat release and low ignition temperature, the front roughness (defined as the RMS deviation of the ignition temperature contour from the average flame position) grew following a power law dependence that was in excellent agreement with the KPZ universality class ($\beta = 1/3$).  As the reaction time was increased, lower values of the roughening exponent were observed, and at a sufficiently great value of reaction time, reversion to a steady, constant-width thermal flame was observed that matched the solution from classical combustion theory.  Deviation away from KPZ scaling was also observed as the ignition temperature was increased.  The features of this system that permit it to exhibit both KPZ and non-KPZ scaling are discussed.
\end{abstract}


\maketitle

\section{Introduction}
\subsection{Flame propagation in random media}
In heterogeneous media with spatially localized, exothermic sources, statistical variations in the concentration of the sources in the media can occur on scales comparable to the thickness of flame fronts propagating through the media. This situation is distinct from homogenous reactive media, as exemplified by gaseous mixtures, where fluctuations in concentration only occur on scales much smaller (e.g., mean molecular spacing) than the flame thickness, permitting a clear separation of scales by several orders of magnitude.  In heterogeneous media (e.g., fuel particulates suspended in gaseous oxidizer \cite{Pokhil1972,Julie2015CF} or mixtures of solid reactants as encountered in Self-propagating High-temperature Synthesis, SHS \cite{Mukasyan2008,Rogachev2015}), the scale of the heterogeneity can be comparable to or greater than the flame thickness, raising the question of how the statistical properties of the media influence the flame dynamics.\\

The fact that a medium has a random, heterogeneous nature does not necessarily imply that the flame will have a rough structure.  Prior studies \cite{GoroshinLeeShoshin1998,Tang2009CTM,Goroshin2011PRE,Tang2011PRE,Mi2016PROCI} have identified a discreteness parameter, consisting of the ratio of the scale of the flame front thickness to the scale of heterogeneity (or, equivalently, the ratio of the reaction time to the diffusion time between the heterogeneities) as the factor that determines if the structure of the flame will manifest the spatial inhomogeneity of the media.  Only in the case where the discreteness parameter is less than unity will the flame itself have a spatial structure determined by the random media.  If the flame thickness is much greater than the spatial scale of the heterogeneity, then the flame may still be accurately treated using a continuum-based description for which the discrete nature of the media only enters via the reaction rate term.  Combustion of a rich suspension of volatile dust in air, for example, can be well described by such a continuum-based model. The case of a dilute suspension of fast burning fuel in an oxidizer, on the other hand, can result in a discreteness parameter less than unity, where the discreteness of the media dominates the propagation dynamics. This effect can be accentuated by using a medium with low thermal diffusivity.  The existence of this regime has been demonstrated experimentally by Goroshin \textit{et al}. \cite{Goroshin2011PRE} and Wright \textit{et al}. \cite{Wright2016}, and has now been recognized as a separate branch of combustion, termed \textit{discrete combustion} covered in review articles \cite{Mukasyan2008,Rogachev2015}.  The efforts to date in examining flame propagation in the discrete combustion regime have focused mostly on the average flame propagation velocity and limits to propagation \cite{Beck2003,Beck20032D,Tang2009CTM,Goroshin2011PRE,Tang2011PRE}, while the structure of the reactive front has not been examined in detail.  In this paper, we will examine the influence of the spatially random media on flame propagation in an idealized system consisting of point-like discrete heat sources.

\subsection{Front roughening in statistical physics}
Over the last three decades, the study of kinetic roughening (distinct from chemical kinetics, \textit{kinetic roughening} refers to fronts becoming increasingly irregular over time, often studied in kinetic processes) of propagating fronts has identified a universality in how the front roughness increases under the influence of stochastic noise.  This line of investigation, which evolved out of non-equilibrium statistical mechanics, has shown that propagation of fronts in a number of seemingly disparate systems (e.g., a bacterial colony spreading across a petri dish, polymerization fronts) are described by a single class of equations that exhibit a power-law growth in the roughness of the front.  More remarkably, a single value of roughening exponent $\beta$ is sufficient to describe the initial growth of a large number of different physical systems regardless of the particular underlying physics. Such systems are said to belong to the same universality class.\\

Systems marked by the presence of stochastic noise, locality of interactions, locally normal front propagation, and front surface tension are said to belong to the Kardar-Parisi-Zhang (KPZ) universality class. Correspondingly, fronts propagating through these systems are described by the KPZ stochastic differential equation \cite{KPZ1986,Halpin1995}; from here, front roughness can be shown through renormalization techniques and verified numerically to increase as a power-law according to the exact exponent $\beta = 1/3$ in two-dimensional media. Many experiments attempting to show KPZ power-law growth have been conducted. One-third power-law front roughening has been found directly in electric field-induced nematic liquid crystal convection \cite{Takeuchi2011} and in electrodeposition \cite{Schilardi1999}. The KPZ universality class has been suggested to provide useful analogues for stromatolite growth \cite{Grotzinger1999} and diversified stock portfolio optimization, where stock capital is treated like front roughness \cite{Marsili1998}. More experiments and models that explored KPZ kinetic roughening can be found in the review by Halpin \textit{et al}.\cite{Halpin1995}\\

Experimental investigation of kinetic roughening in combustion has been, to our knowledge, limited to investigations of paper smoldering \cite{Zhang1992,Maunuksela1997,Myllys2001}. In these studies, the front roughening likely reflects the inhomogeneous microstructure of the smoldering paper, but not the inherent flame structure. Numerical simulations of flame propagation in reactive media with randomly distributed sources governed by Arrhenius kinetics were performed by Provatas \textit{et al}. \cite{Provatas1995,Provatas1995Scaling}, demonstrating that the kinetic roughening of the flame in their system belongs to the KPZ universality class. The conditions under which their model would result in a smooth flame front, reverting to the flame structure described by classical, continuum-based combustion theory, was not explored.\\ 

The purpose of this paper is to investigate the front structure in a model system proposed by Goroshin \textit{et al}. \cite{GoroshinLeeShoshin1998} and Beck and Volpert \cite{Beck2003,Beck20032D} (see Sec~\ref{Sec2_1}), and to quantify the structure, and specifically the roughening, of the front. The asymptotic limits of both highly discrete and (at the other extreme) continuum-like behavior will be examined, as well as the spectrum of behavior in between, using both classical thermal flame theory and the tools of kinetic front roughening to provide a formalism to assist in the interpretation of the results. The possibility that the model system used in this paper may fall into the class of KPZ universality will also be explored.

\section{Model}
\label{Sec2}
\subsection{Formulation}
\label{Sec2_1}
The governing equation of this model is a two-dimensional heat diffusion equation with a source term that describes the heat release of point-like, randomly distributed reactive particles. The dimensional form of the governing equation is formulated as follows (note that tilde ``$\sim$'' indicates dimensional quantities),
\begin{equation}
\frac{\partial \widetilde{T}}{\partial \widetilde{t}} = \widetilde{\alpha} \left( \frac{\partial^2 \widetilde{T}}{\partial \widetilde{x}^2} + \frac{\partial^2 \widetilde{T}}{\partial \widetilde{y}^2} \right) + \frac{\widetilde{q}\widetilde{B}}{\widetilde{\rho}\,\widetilde{c}_{p}} \widetilde{R}
\label{Eq1}
\end{equation}
where $\widetilde{\alpha}$, $\widetilde{\rho}$, and $\widetilde{c}_{p}$ denote the thermal diffusivity, density, and specific heat capacity of the medium in which the point sources are distributed, respectively. For simplicity, $\widetilde{\alpha}$, $\widetilde{\rho}$, and $\widetilde{c}_{p}$ are assumed to be constant in this model. The assumption of constant properties is not as nonphysical as might be assumed, because of the fact that an increase in thermal diffusivity of the medium at higher temperature approximately compensates the effect of volumetric dilation (i.e., decrease in density) due to heat release. The heat release per unit mass of the fuel is denoted as $\widetilde{q}$. The fuel mass per unit volume, $\widetilde{B}$, can be expressed as the total number of reactive particles $N$ multiplied by the mass of each individual particle $\widetilde{m}_\mathrm{p}$ and divided by the volume (area) of the two-dimensional system $\widetilde{V}_\mathrm{sys}$, i.e., $\widetilde{B}=N \widetilde{m}_\mathrm{p}/\widetilde{V}_\mathrm{sys}$. The average spacing between two neighboring particles $\widetilde{l}$ can be expressed as $\widetilde{l}=\sqrt{\widetilde{V}_\mathrm{sys}/N}$.\\

The overall rate of heat release of the spatially discrete reacting particles in the system is represented by $\widetilde{R}$ in Eq.~\ref{Eq1}. Since these particles are considered as point sources in this model, i.e., they occupy an infinitely small volume, the location where each (the $i^\mathrm{th}$) particle releases heat can be described in $\widetilde{R}$ as a spatially two-dimensional $\mathrm{\delta}$-function, $\delta \left(\widetilde{x}-\widetilde{x}_i \right) \delta \left(\widetilde{y}-\widetilde{y}_i \right)$. Once the temperature of a source reaches the prescribed ignition temperature $\widetilde{T}_\mathrm{ign}$, the source begins to release its heat. The heat release process can occur instantaneously ($\widetilde{t}_\mathrm{r}=0$) or over a finite amount of time ($\widetilde{t}_\mathrm{r}>0$). For the case of $\widetilde{t}_\mathrm{r}=0$, the rate of heat release is a temporal  $\delta$-function; for $\widetilde{t}_\mathrm{r}>0$, the rate of heat release is a temporal boxcar function that is constructed from two Heaviside functions. Considering the spatial discreteness of the randomly distributed sources with either zero or a finite heat release time, the rate of heat release of the entire system, $\widetilde{R}$, can be obtained as a linear superposition (summation) of all $N$ heat sources and expressed as follows,
\begin{equation}
\widetilde{R} = 
\begin{cases}
\sum_{i=1}^{N} \delta \left(\widetilde{x}-\widetilde{x}_i \right) \delta \left(\widetilde{y}-\widetilde{y}_i \right) \delta \left(\widetilde{t}-\widetilde{t}_{\mathrm{ign},i} \right) \;\;\; \mathrm{for} \;\;\; \widetilde{t}_\mathrm{r}=0\\
\sum_{i=1}^{N} \frac{1}{\widetilde{t}_\mathrm{r}} \delta \left(\widetilde{x}-\widetilde{x}_i \right) \delta \left(\widetilde{y}-\widetilde{y}_i \right) \mathrm{H} \left(\widetilde{t}-\widetilde{t}_{\mathrm{ign},i} \right) \mathrm{H} \left(\widetilde{t}_\mathrm{r}-\widetilde{t}+\widetilde{t}_{\mathrm{ign},i} \right) \;\;\; \mathrm{for} \;\;\; \widetilde{t}_\mathrm{r}>0
\end{cases}
\label{Eq2}
\end{equation}
where $\widetilde{t}_{\mathrm{ign},i}$ is the time that has elapsed since $\widetilde{t}=0$ until the temperature of the $i^\mathrm{th}$ source first reaches $\widetilde{T}_\mathrm{ign}$.\\

The governing equation used for calculation is in a dimensionless form. The spatial coordinates are normalized by the average particles spacing $\widetilde{l}$, i.e., $x=\widetilde{x}/\widetilde{l}$ and $y=\widetilde{y}/\widetilde{l}$. Hence, the dimensionless particle spacing is unity, i.e., $l=1$. Time is nondimensionalized by the characteristic time scale of heat diffusion between particles, $t=\widetilde{t}/\widetilde{t}_\mathrm{d}$ where $\widetilde{t}_\mathrm{d}=\widetilde{l}^2/\widetilde{\alpha}$. The importance of source discreteness can be quantified by comparing the heat release time of each particle $\widetilde{t}_\mathrm{r}$ to the interparticle diffusion time $\widetilde{t}_\mathrm{d}$, yielding the dimensionless discreteness parameter $\tau_\mathrm{c}$
\begin{equation}
\tau_\mathrm{c} = \frac{\widetilde{t}_\mathrm{r}}{\widetilde{t}_\mathrm{d}} = \frac{\widetilde{\alpha} \widetilde{t}_\mathrm{r}}{\widetilde{l}^2} 
\label{Eq3}
\end{equation}
When $\tau_\mathrm{c} \gg 1$, continuum flame behavior is recovered as inter-particle heat diffusion times become insignificant; conversely, when $\tau_\mathrm{c} \ll 1$, the discreteness of the heat sources strongly influences flame propagation. In the discrete limit of $\tau_\mathrm{c} = 0$, heat is released instantaneously.\\

The temperature $\theta$ of the system is nondimensionalized as
\begin{equation}
\theta = \frac{\widetilde{T}-\widetilde{T}_0}{\widetilde{T}_\mathrm{ad}-\widetilde{T}_0}
\label{Eq4}
\end{equation}
where $\widetilde{T}_0$ is the initial temperature of the medium and $\widetilde{T}_\mathrm{ad}$ is the adiabatic flame temperature, i.e., $\widetilde{T}_\mathrm{ad}=\widetilde{T}_0+{\widetilde{q}\widetilde{B}}/\left({\widetilde{\rho}\,\widetilde{c}_{p}}\right)$. Thus, $\theta=0$ corresponds to the initial temperature, and $\theta=1$ corresponds to the adiabatic flame temperature. Continuum flame theory suggests that flame propagation is limited to dimensionless ignition temperatures less than unity, i.e., $\theta_\mathrm{ign}<1$.\\

\subsection{Solution}
The governing equation is the linear heat equation (Eq.~\ref{Eq1}) with a source term consisting of spatial $\delta$-functions (point sources), given by Eq.~\ref{Eq2}. The solution of the temperature at a specific location in this system, $\theta(x,y,t)$, can be obtained via the contribution from all ignited point sources. This solution can be constructed analytically via linear superposition of the Green's function solution of a single source for $\tau_\mathrm{c} = 0$, or its time-convolution with a prescribed heat release profile (i.e., a temporal boxcar function described by Eq.~\ref{Eq2}) for $\tau_\mathrm{c} > 0$ as follows, respectively
\begin{equation}
\theta(x,y,t)= 
\begin{cases}
\mathlarger{‎‎\sum}_{i=1}^{N}  \frac{\mathrm{H}\left(t-t_{\mathrm{ign},i} \right)}{4\pi \left(t-t_{\mathrm{ign},i} \right)} \exp{\left(-\frac{\left(x-x_i\right)^2 + \left(y-y_i\right)^2}{4 \left(t-t_{\mathrm{ign},i} \right)}\right)} \;\;\; \mathrm{for} \;\;\; \tau_\mathrm{c}=0\\[0.5cm]
\mathlarger{‎‎\sum}_{i=1}^{N} \frac{\mathrm{H}\left(t-t_{\mathrm{ign},i} \right)}{\tau_\mathrm{c}} \bigintsss_{\left(t-t_{\mathrm{ign},i}- \tau_\mathrm{c}\right)\mathrm{H}\left(t-t_{\mathrm{ign},i}- \tau_\mathrm{c}\right)}^{t-t_{\mathrm{ign},i}} \mathlarger{\frac{1}{4\pi\zeta}} \exp{\left(-\frac{\left(x-x_i\right)^2 +\left(y-y_i\right)^2}{4\zeta} \right)}  \mathrm{d}\zeta \;\;\; \mathrm{for} \;\;\; \tau_\mathrm{c}>0
\end{cases}
\label{Eq5}
\end{equation}
where $\zeta$ is a dummy variable for time $t$ in the integral. Recall that $\left(x_i,y_i \right)$ is the position of the $i^\mathrm{th}$ source, and $t_{\mathrm{ign},i}$ is the time when the temperature of the $i^\mathrm{th}$ source first reaches $\theta_\mathrm{ign}$. Thus, the two parameters $\tau_\mathrm{c}$ and $\theta_\mathrm{ign}$ completely characterize the solution for this system of discrete sources. The complete details of the model and its analytic solutions can be found in \cite{Tang2009CTM,Tang2011PRE}, along with studies of the average flame speed and limits to propagation.

\subsection{Implementation}
For most of the simulations performed in this study, two-dimensional, square domains were initialized with $40000$ point sources randomly distributed (domain size of $200$ by $200$ average particle spacing) using the MT19937 pseudorandom number generator \cite{Matsumoto1998}. For each $\tau_\mathrm{c}$ and $\theta_\mathrm{ign}$, a random distribution of sources was generated, and the flame was initiated by forcing $10\%$ of the domain's sources at one (left) end of the domain to ignite simultaneously at the beginning of the simulation. Thus, in all cases of simulations presented in this paper, the flame front propagates rightward. Periodic boundary conditions were applied to the edges of the domain parallel to the direction of flame propagation by considering images (copies) of the domain placed on either side of the original. Considering more than one image on each side was found to have no influence on the flame for the chosen domain size. The other edges (left and right) were left with open boundary conditions, allowing heat to diffuse out of the domain. Each randomly generated domain occupied a single HPC (high-performance computing) core for computing the solution of the temperature field and subsequent post-processing. The simulations for each set of $\theta_\mathrm{ign}$ and $\tau_\mathrm{c}$ were performed $120$ times  over different clusters on four supercomputer servers, with wall-clock times ranging from several hours (near $\tau_\mathrm{c}=0$) up to several weeks (high $\tau_\mathrm{c}$).\\
\begin{figure}
\centerline{\includegraphics[width=0.7\textwidth]{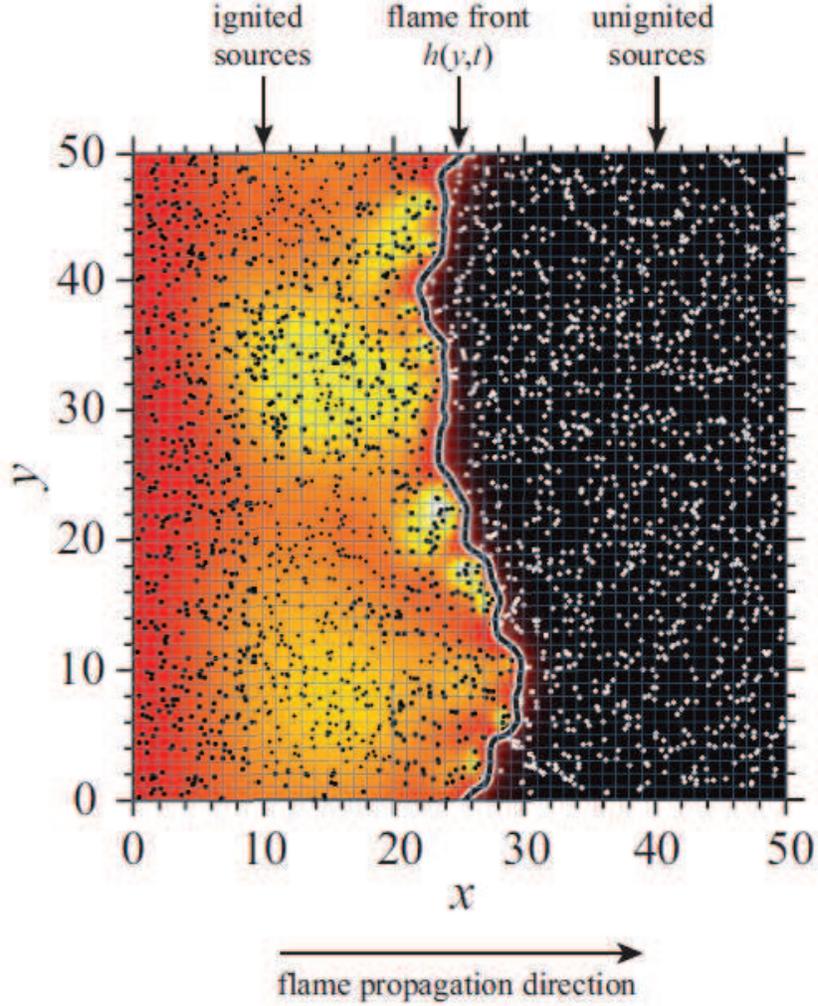}}
\caption{Visualization of a small domain ($50\times50$) with $\theta_\mathrm{ign} = 0.5$, $\tau_\mathrm{c} = 1$, showing the color contours of temperature field, with the flame front (marked by the black-white line), and randomly distributed ignited (black dots) and unignited (white dots) sources. The uniform Cartesian grid over which the temperature field is calculated for front roughness measurement is illustrated as the blue lines.}
\label{Fig1}
\end{figure}

For each run, via solving the temperature at the location of each source $\left(x_i,y_i \right)$, the ignition time of each source $t_{\mathrm{ign},i}$ was determined one after another in an event-driven manner. Once the solution was completed, post-processing of the solution consisted of calculating the temperature field $\theta(x,y,t)$ at specific times by evaluating Eq.~\ref{Eq5} on a uniform Cartesian grid over the entire domain with the obtained sequence of ignition times. A sample snapshot of the resulting temperature field for the case with $\theta_\mathrm{ign}=0.5$ and  $\tau_\mathrm{c}=1$ is shown as a color contour plot in Fig.~\ref{Fig1}. The color levels are linearly scaled between $\theta = 0$ (dark) and $1.5$ (bright). Ignited and unignited point sources are marked as black and white dots, respectively. Only the ignited sources contribute to the temperature field. The rightward-propagating flame front is highlighted by the black-white line. Each simulation was ended once the furthest (rightmost) point along the flame front reached the right end of the domain. The blue lines in Fig.~\ref{Fig1} illustrate the uniform grid used to calculate the temperature field for post-processing analysis of the front roughness. The resolution of the grid was verified to be sufficient so that further refining it would not influence the resulting flame front position and roughness. For illustrative clarity, however, the grid plotted as blue lines in Fig.~\ref{Fig1} is more sparse than that actually used to calculate the temperature field. The detection of the flame front and the measurement of its roughness on each snapshot of the temperature field will be described in Sec.~\ref{Sec4_1} and \ref{Sec4_2}, respectively.

\section{Results}
\begin{figure}
\centerline{\includegraphics[width=1.0\textwidth]{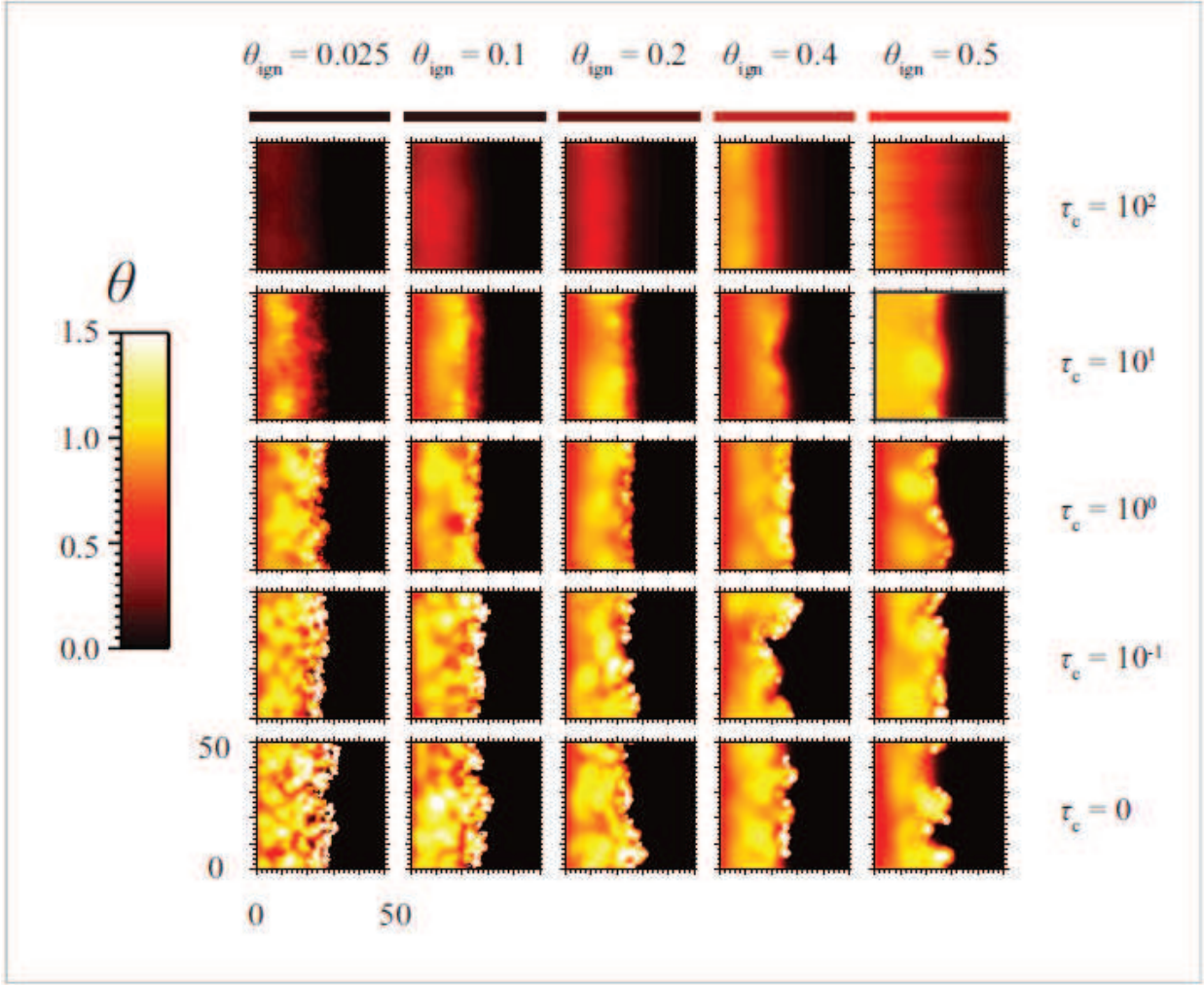}}
\caption{Temperature fields of flames propagating rightward in systems of discrete sources. Average inter-source spacing is unity, and size of each plot window is chosen to be $50\times50$. Each system corresponds to a different set of nondimensional ignition temperature $\theta_\mathrm{ign}$ (increasing from left to right) and discreteness parameter $\tau_\mathrm{c}$ (increasing from bottom to top). The color bars above the top row correspond to the column's ignition temperature. The flame front is near the middle of each plot window.}
\label{Fig2}
\end{figure}

Snapshots of the resulting temperature fields for a variety of parameters $\theta_\mathrm{ign}$ and $\tau_\mathrm{c}$ are shown as color contour plots in Fig.~\ref{Fig2}. In this study, $\theta_\mathrm{ign}$ was varied from $0.025$ to $0.5$; values of $\theta_\mathrm{ign}$ exceeding $0.5$ exhibit complex dynamics leading to quenching.\cite{Tang2009CTM,Tang2011PRE}  Values of $\tau_\mathrm{c}$ examined included zero and finite values ranging from $10^{-2}$ to $10^2$.The color levels are linearly scaled between $\theta = 0$ and $1.5$. The snapshots in Fig. 2 are arranged in an array with $\theta_\mathrm{ign}$ increasing from left to right and $\tau_\mathrm{c}$ decreasing from top to bottom.  In each snapshot, the flame propagates rightward. The flame front (iso-contour of ignition temperature) is roughly in the middle of each plot. Although each case of computation was performed over a domain of $40000$ sources, the snapshots shown in Fig. 2 are plotted over a zoomed-in area that contains approximately $2500$ sources in order to clearly show the features of the flame front.\\

For large values of $\tau_\mathrm{c}$ (top rows in Fig.~\ref{Fig2}), the flame has a smooth and planar structure. The resulting temperature fields are nearly transversely (i.e., in the $y$-direction that is perpendicular to flame propagation direction) uniform. The right-to-left (i.e., from before to after the heat release) span of the region associated with a gradual increase in temperature is nearly half of the plot window size ($50$ units of length), much longer than the average source spacing (one unit of length), especially in the cases with large $\theta_\mathrm{ign}$.  For small values of $\tau_\mathrm{c}$ (bottom rows in Fig.~\ref{Fig2}), the resulting flame fronts are transversely irregular and featured with randomly located hot (bright) and cold (dark) spots and peninsulas with a characteristic length that is comparable to the average source spacing.\\

\begin{figure}
\centerline{\includegraphics[width=1.0\textwidth]{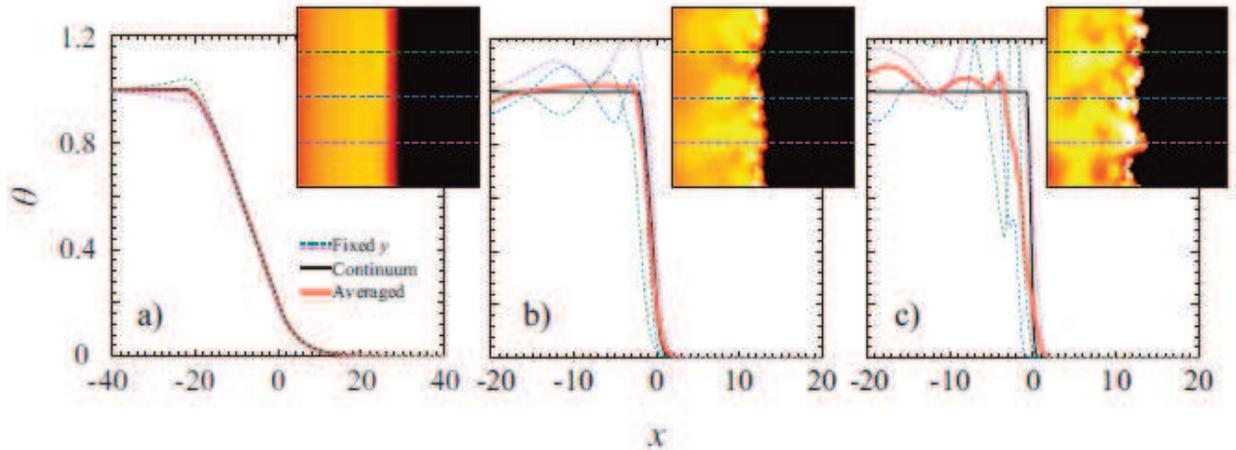}}
\caption{Temperature profiles plotted against spatial coordinate in direction of flame propagation, i.e., $\theta(x)$, for ignition temperature $\theta_\mathrm{ign} = 0.2$ and discreteness parameter (a) $\tau_\mathrm{c} = 100$, (b) $\tau_\mathrm{c} = 1$, (c) $\tau_\mathrm{c} = 0.1$. Corner inset: corresponding temperature contour plots. Thin dashed curves in temperature profile are extracted from corresponding lines shown on the temperature contours. Red line corresponds to width-averaged temperature profile, and black line corresponds to the analytic solution based on a continuum flame model.}
\label{Fig3}
\end{figure}

In Fig.~\ref{Fig3}, the width-averaged temperature profiles (solid red curves) along the direction of propagation for three cases with the same ignition temperature $\theta_\mathrm{ign} = 0.2$, and different values for $\tau_\mathrm{c}$, i.e., $\tau_\mathrm{c} = 100 \gg 1$, $\tau_\mathrm{c}  = 1$, and $\tau_\mathrm{c} = 0.1 \ll 1$, are shown and compared with the analytic solution of a thermal flame (solid black curve) based on the assumption of a continuous reactive medium. Each plot of temperature profile is accompanied by the corresponding contour plot of the temperature field. Temperature profiles extracted from the temperature field along the corresponding thin dotted lines are also plotted in Fig.~\ref{Fig3}.\\

For the case of $\tau_\mathrm{c} = 100$ shown in Fig.~\ref{Fig3}(a), the averaged temperature profile and the temperature profiles extracted from the resulting temperature field are nearly indistinguishable from the continuum solution. (The details of the continuum solution can be found in Ref.~\cite{Tang2011thesis}.) As shown in Fig.~\ref{Fig3}(b), for the case of $\tau_\mathrm{c} = 1$, the averaged temperature profile has a steeper slope, and lies fairly close to the continuum solution. However, the extracted temperature profiles in this case exhibit large fluctuations around the averaged profile and the continuum solution. Figure~\ref{Fig3}(c) shows the case with $\tau_\mathrm{c}  = 0.1$, where both averaged and extracted temperature profiles exhibit large fluctuations and significantly deviate from the continuum solution.

\section{Analysis}
\label{Sec4}
After obtaining the ignition time $t_{\mathrm{ign},i}$ of each source and the complete time history of the temperature field $\theta(x,y,t)$ from the simulation, the flame front position $h(y,t)$ was detected and recorded over time. Based on the data of $h(y,t)$ for a statistical ensemble with a specific set of $\theta_\mathrm{ign}$ and $\tau_\mathrm{c}$, the time history of the ensemble-averaged front roughness $W(t)$ was calculated. Whether the obtained $W(t)$ exhibits a power-law growth over time, i.e., $W \sim t^\beta$, was then examined, and for such cases, the roughening exponent $\beta$ was determined. Each step of the above-described analysis is elaborated in the following subsections.
   
\subsection{Detection of flame front position $h$}
\label{Sec4_1}
With $\theta(x, y)$ obtained at time $t$ on a uniform Cartesian grid as illustrated in Fig.~\ref{Fig1}, the position where temperature first reached $\theta_\mathrm{ign}$ was designated as the flame front location and was found by searching leftward from the right end along each horizontal ($x$-direction) array of grid points at a given $y$-coordinate. By performing this search across the width of the domain, flame front position as a function of $y$ and $t$, i.e., $h(y,t)$, was thus obtained. The width-averaged (over the $y$-direction) position of the flame front as a function of $t$, i.e., $\overline{h}(t)$, was then calculated as follows,
\begin{equation}
\label{Eq6}
\overline{h}(t) = \frac{1}{L} \int_{0}^{L} h(y,t) \mathrm{d}y
\end{equation}
where $L$ is the width of the domain in the $y$-direction, which is $200$ units of length in all the simulations reported in this paper. Note that the integration in Eq.~\ref{Eq6} was calculated numerically over the same Cartesian grid used for post-processing temperature field.

\subsection{Measurement of front roughness $W$}
\label{Sec4_2}
The flame front roughness can be quantified as the root mean square of the difference between flame front position $h$ and the width-averaged (over $y$-direction) front position $\overline{h}$, averaged over a statistical ensemble, 
\begin{equation}
W(t) = {\left\langle \sqrt{\frac{1}{L} \int_{0}^{L} \left(h(y,t)-\overline{h}(t) \right)^2 \mathrm{d}y} \right\rangle}
\label{Eq7}
\end{equation}
where ${\langle~\rangle}$ denotes averaging over a statistical ensemble (of $120$ independent simulations in this present study).\\

\subsection{Measurement of roughening exponent $\beta$}
\label{Sec4_3}
\begin{figure}
\centerline{\includegraphics[width=1.0\textwidth]{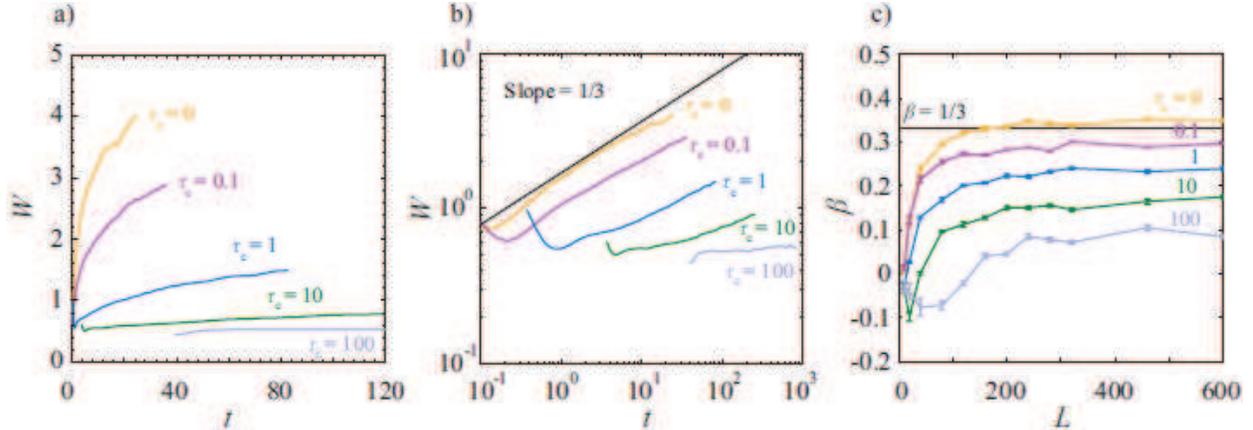}}
\caption{Ensemble-averaged flame front roughness $W$ as a function of time $t$ for $\theta_\mathrm{ign} = 0.2$ and $\tau_\mathrm{c}$ from $0$ to $100$ using (a) linear scales, and (b) logarithmic scales. Each curve is plotted for flame propagation in a $200 \times 200$ domain ($40000$ discrete sources, beginning from the time at which at least $20$ sources have been triggered). The measured $\beta$ is plotted as a function of domain width $L$ in (c). The reference line in (b) and (c) shows $\beta=1/3$. Error bars show the $95\%$ confidence interval on the slope measurement.}
\label{Fig4}
\end{figure}

The ensemble-averaged front roughness $W$ over time $t$ for an example case of $\theta_\mathrm{ign} = 0.20$ is shown in Fig.~\ref{Fig4}(a) for values of $\tau_\mathrm{c}$ varied over several orders of magnitude. Due to the initiation method, the flame front begins with a small value of roughness.  Overall front roughness decreases as $\tau_\mathrm{c}$ is increased; see Fig.~\ref{Fig2} for example.\\

Figure~\ref{Fig4}(b) shows the ensemble-averaged flame front roughness $W$ as a function of time $t$ plotted using logarithmic scales for $\theta_\mathrm{ign} = 0.20$ and various $\tau_\mathrm{c}$. If the front roughness exhibits a power-law growth over time, the log-log plot of $W(t)$ must appear as a positively sloped straight line. Hence, a linear fit in the logarithmic domain, after removal of large initiation transients, was used to measure $\beta$ for each set of $\tau_\mathrm{c}$ and $\theta_\mathrm{ign}$. At low $\tau_\mathrm{c}$, a good power-law agreement is found, with slope (power-law exponent) $\beta$ very near $1/3$. As $\tau_\mathrm{c}$ is increased, the slope decreases, although the log-log plot exhibits linear behavior for $\tau_\mathrm{c}$ up to unity. A fit through the data yields $\beta = 0.33$ for $\tau_\mathrm{c} = 0$, $\beta = 0.28$ for $\tau_\mathrm{c} = 0.1$, and $\beta = 0.26$ for $\tau_\mathrm{c} = 1$. As $\tau_\mathrm{c}$ is increased far beyond unity up to $\tau_\mathrm{c} = 100$, the ensemble-averaged front roughness $W$ becomes much smaller than that resulting from the cases with $\tau_\mathrm{c} \ll 1$. The slope identified in the log-log plots for the cases with large $\tau_\mathrm{c}$ is nearly zero.\\

In order to investigate the effect of the width of the computational domain on the results, for each case plotted in Fig.~\ref{Fig4}(a) and (b), an ensemble of simulations was also performed using various domain widths ranging from $L=10$ to $600$ for a domain length fixed at $200$. The roughening exponent $\beta$ measured from the ensemble-averaged results of $W(t)$ is plotted as functions of $L$ in Fig.~\ref{Fig4}(c). The measured $\beta$ shown in Fig.~\ref{Fig4}(c) gradually increases and approaches a plateau value as $L$ increases, and exhibits no significant change beyond $L=200$.\\
\begin{figure}
\centerline{\includegraphics[width=0.7\textwidth]{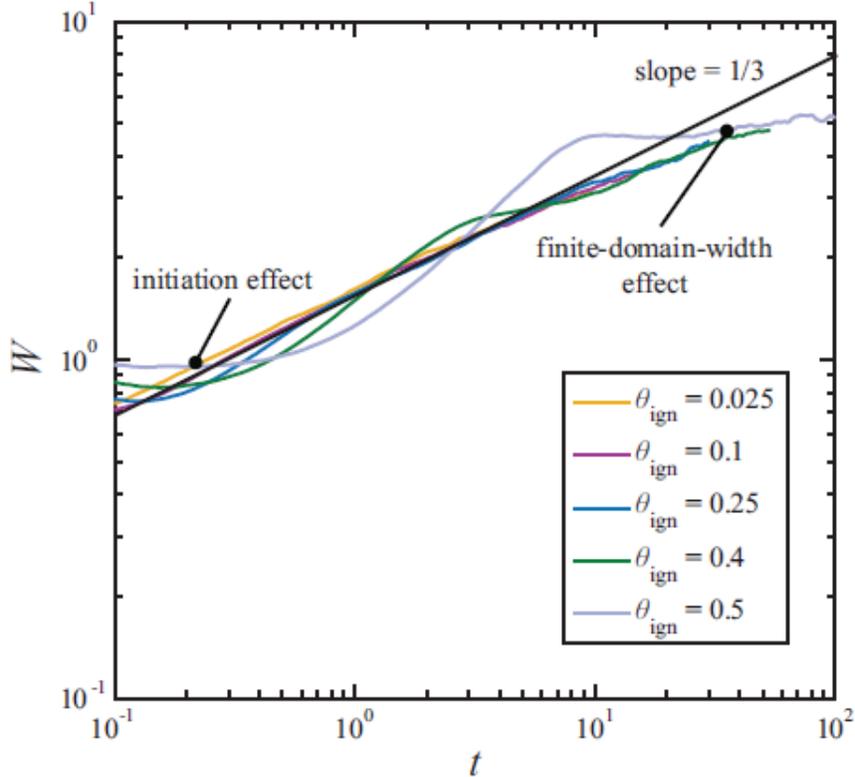}}
\caption{Ensemble-averaged flame front roughness $W$ as a function of time $t$ for $\tau_\mathrm{c}=0$ and $\theta_\mathrm{ign} = 0.025$ to $0.5$ using logarithmic scales. Each curve is plotted for flame propagation in a $200 \times 200$ domain ($40000$ discrete sources). The roughening behavior influenced by the initiation of the flame and the finite domain width are indicated on the curve of $\theta_\mathrm{ign} = 0.5$.}
\label{Fig5}
\end{figure}

Figure~\ref{Fig5} shows the ensemble-averaged flame front roughness $W$ against time $t$ using logarithmic scales for the cases of $\tau_\mathrm{c} = 0$ and various values of $\theta_\mathrm{ign}$. The curves associated with $\theta_\mathrm{ign} =0.025$ and $0.1$ appear to be fairly straight lines in the log-log plot. For these cases, the roughening exponent $\beta$ was measured consistently to be near $\beta = 0.33\pm0.02$. As $\theta_\mathrm{ign}$ increases from $0.25$ to $0.5$, the corresponding curve of $W(t)$ on log-log scales becomes increasingly S-shaped, i.e., with two flattened parts at early and late times. For the cases associated with an S-shaped curve of $W(t)$ that significantly deviates from a straight line, the resulting roughening behavior is not considered as power-law growth.\\
\begin{figure}
\centerline{\includegraphics[width=1.0\textwidth]{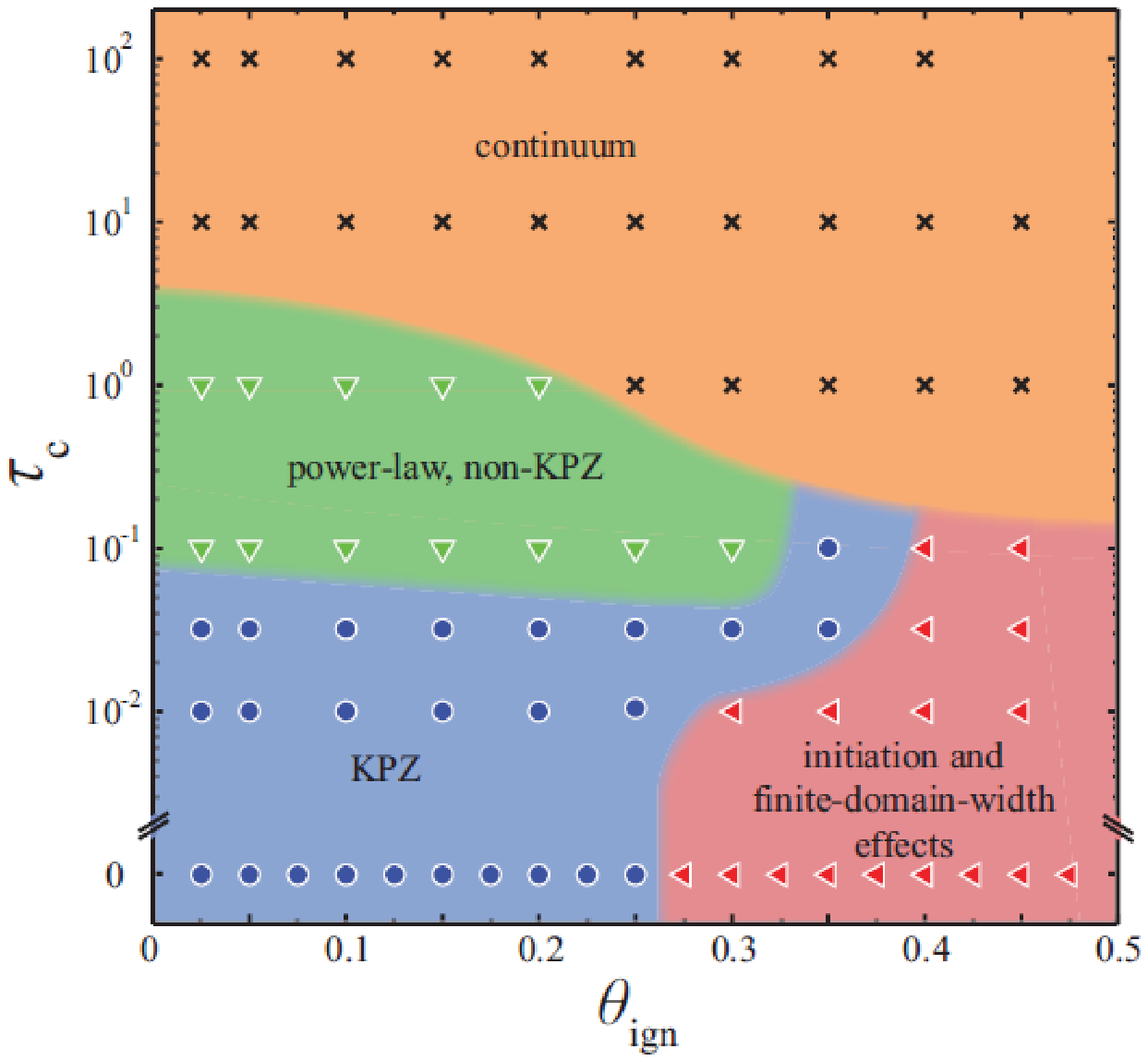}}
\caption{Simulation results categorized according to the flame front roughening behavior and summarized as a map in $\theta_\mathrm{ign}$-$\tau_\mathrm{c}$ parameter space. Each color-shaded region represents a category of front roughening behavior; each symbol represents an ensemble of simulations with a specific set of $\theta_\mathrm{ign}$ and $\tau_\mathrm{c}$.}
\label{Fig6}
\end{figure}

From the analysis of the simulation results, four different categories of flame front roughening have been identified and summarized in $\theta_\mathrm{ign}$--$\tau_\mathrm{c}$ parameter space (Fig.~\ref{Fig6}) as follows: 1. power-law roughening with an exponent $\beta \simeq 1/3$, or KPZ kinetic roughening (blue-shaded region with circles); 2. power-law roughening with an exponent $0<\beta<1/3$ (green-shaded region with downward-pointing triangles); 3. non-power-law roughening (red-shaded region with left-pointing triangles); 4. non-roughening flame front with $\beta \simeq 0$ (orange-shaded region with crosses). Note that, in Fig.~\ref{Fig6}, the boundaries of the color-shaded regions are defined as where the ensemble-averaged front roughness growth deviated by 0.04 (above or below) the KPZ value of $\beta = 1/3$ or yielded a coefficient of determination $r^2 < 0.99$. Thus, there is some degree of arbitrariness in defining these regions, but the different classifications proposed here represent qualitatively different regimes of front behavior.

\section{Discussion}
As shown in Figs.~\ref{Fig4} and~\ref{Fig5}, for $\theta_\mathrm{ign} \lesssim 0.25$ and $\tau_\mathrm{c} \ll 1$ (summarized in Fig.~\ref{Fig6} as blue circles), the growth of flame front roughness follows a power law with an exponent $\beta \simeq 0.33$ that is in good agreement with the one-third exact exponent predicted by KPZ kinetic roughening theory. With a short reaction time (small $\tau_\mathrm{c}$) and a small increase in temperature required for each source to be ignited (small $\theta_\mathrm{ign}$), the heat release of randomly distributed sources only locally interact with the propagating flame front, while heat diffusion through the medium always has a smoothing effect on the flame front. This physical mechanism is well described by the nonlinear KPZ interface equation. Hence, the resulting front roughening from the cases with small $\theta_\mathrm{ign}$ and $\tau_\mathrm{c}$ clearly belongs to the KPZ universality class.\\

For the cases with $\tau_\mathrm{c} > 0.1$ or $\theta_\mathrm{ign} > 0.25$, The growth of flame front roughness deviates from KPZ kinetic roughening as shown in Figs.~\ref{Fig4} and~\ref{Fig5}, respectively. With a higher $\theta_\mathrm{ign}$, a source needs to absorb the heat released by more ignited sources from a larger area; with a larger $\tau_\mathrm{c}$, i.e., longer reaction time, the heat released by a source can diffuse far away from its location. Both of these effects render the interaction between the heat release of sources and the propagation of the flame front less dependent on the local sources exclusively, and the global diffusion of heat from all prior sources becomes the dominant mechanism of propagation. Since the locality of the interface growth mechanism is one of the necessary conditions for KPZ kinetic roughening to occur, increasing $\theta_\mathrm{ign}$ or $\tau_\mathrm{c}$ in this system leads to non-KPZ roughening behavior (i.e., moving from the blue-shaded KPZ region into the red- or green-shaded non-KPZ regions, respectively, as shown in Fig.~\ref{Fig6}). In the following paragraphs, the discussion will be focused on the categories of non-KPZ behavior identified in this study.\\

Although the non-locality of the underlying propagation mechanism at high $\theta_\mathrm{ign}$ is believed to result in an overall non-KPZ roughening behavior, the S-shaped curves of $W(t)$ in log-log scales for the cases with $\theta_\mathrm{ign} \gtrsim 0.25$ (shown in Fig.~\ref{Fig5}) are mainly attributable to the initiation method used in the simulations and a finite-domain-width effect. Since the flame propagation speed is slower in the cases with higher $\theta_\mathrm{ign}$, the transient process of front roughening after the initiation, i.e., the initial flattened part of the $W(t)$ curves as indicated on the curve of $\theta_\mathrm{ign}=0.5$ in Fig.~\ref{Fig5}, persists longer compared to those in the cases with lower $\theta_\mathrm{ign}$. Thus, as this long initiation-influenced process ends, the flame front almost enters the region influenced by the finite width of the domain (with periodic boundaries), the late-time flattened part of the $W(t)$ curves as indicated on the curve of $\theta_\mathrm{ign}=0.5$ in Fig.~\ref{Fig5}. Using a larger (longer and wider) simulation domain would likely reduce the initiation and finite-domain-width effects on the roughening behavior. Given a very low flame propagation speed at high $\theta_\mathrm{ign}$, however, performing these simulations in a much larger domain would be excessively time-consuming with simulations requiring months of wall-clock time.\\

If $\theta_\mathrm{ign}$ is fixed at a rather small value, which results in a greater flame propagation speed, it is computationally feasible to more clearly explore the transition from KPZ to non-KPZ behavior while verifying its independence on the domain width as $\tau_\mathrm{c}$ varies over orders of magnitude. As shown in Fig.~\ref{Fig4}(b), the slope of the log-log plot of $W(t)$, i.e., the roughening exponent $\beta$, gradually decreases from $1/3$ as $\tau_\mathrm{c}$ increases from $0$ to $100$ with $\theta_\mathrm{ign}=0.2$. One may argue that the non-local front propagation mechanism due to a longer reaction time would become relatively local again as the domain size increases, hence, the roughening behavior would revert back to KPZ kinetic roughening. Indeed, it is well established for the KPZ universality class that when a finite width of domain is used, the front roughness $W$ approaches a limiting value that scales as $L^\chi$, where $\chi = 1/2$ in two-dimensional media \cite{KPZ1986}. The study of domain width reported in Fig.~\ref{Fig4}(c) was performed in order to rule-out this finite-domain-width effect on the present results. As shown in Fig.~\ref{Fig4}(c), it has been verified that the resulting non-KPZ roughening behavior, i.e., with $\beta$ significantly smaller than $1/3$, persists as the transverse width of the domain is increased beyond $L=200$. These results can be explained by considering the fact that the spatial randomness of the reactive medium is of a fixed characteristic length, i.e., the average source spacing of unity; there is no stochastic pattern of source distribution at scales significantly larger than this characteristic length. Thus, the cases showing non-KPZ roughening behavior do not revert back to KPZ roughening when they were simulated in a much wider domain.\\

As the reaction time further increases beyond the characteristic time scale of heat diffusion between neighboring sources, i.e., $\tau_\mathrm{c} \gg 1$ (the orange-shaded region in Fig.~\ref{Fig6}), the spatial heterogeneities imposed by the random distribution of heat sources would be homogenized via heat diffusion over this long duration of heat release. Flame front roughness, as a direct result of spatial heterogeneities in the reactive medium, can thus hardly be developed in the cases with $\tau_\mathrm{c} \gg 1$. For $\tau_\mathrm{c}=100$, the resulting temperature fields appear to be laminar-like, with a planar flame front as shown in the top row of Fig. 2. In Fig. 3(a), the averaged temperature profile across the flame structure for the case with $\tau_\mathrm{c}=100$ closely matches the corresponding continuum-based, laminar flame solution.\\

Although KPZ front roughening has been demonstrated by Provatas \textit{et al.} \cite{Provatas1995,Provatas1995Scaling} in simulations of a combustible medium with randomly distributed reactants, the significance of this current study is that a full spectrum of flame structure and roughening behavior, varying from the laminar flame solution predicted by continuum-based combustion theory to the KPZ kinetic roughening governed by statistical physics, has been explored. Experimental efforts are underway to observe flame propagation in equivalent three dimensional systems using particulates of nonvolatile fuel as lean suspensions in gaseous oxidizers with low thermal diffusivity.\cite{Goroshin2011PRE,Wright2016}

\section{Conclusion}

In the present study, the roughening of flame fronts propagating in a two-dimensional system consisting of point-like heat sources that are activated by the heat diffusing from previously initiated sources was investigated.  The system was amenable to analytic solution, although implementation of the analytic solution for systems with a large number of particles and ensemble averaging of these simulations necessitated use of high-performance computing resources.  For sufficiently short heat release times (as compared to interparticle diffusion time) and low ignition temperatures, a power law growth of the front roughness was obtained in excellent agreement with the predictions of the KPZ universality class ($\beta = 1/3$).  As the heat release time was increased to a time comparable to the interparticle diffusion time, the front continued to exhibit power law growth, but with an exponent less than the KPZ value, i.e., $\beta < 1/3$.  As the heat release time became several orders of magnitude greater than the interparticle diffusion time, no front roughening was observed and a steady flame structure in agreement with classical flame theory was obtained.  The non-KPZ behavior was not attributed to finite domain size effects, as the result persisted as the domain size was increased, but rather to the nonlocality of source interaction as the scale of flame thickness becomes large compared to the scale of the source spacing.  To our knowledge, this is the first time a transition from KPZ behavior to classical, thermal front propagation was demonstrated in a system of experimental relevance.  Failure of KPZ behavior as the source ignition temperature was increased was also noted, although the effect of the domain size could not be definitively ruled out in this case.  A qualitative map of the different regimes of behavior and their dependence upon the reaction time and ignition temperature was proposed, and will be used as a guide in future experimental studies.

\begin{acknowledgments}
This project was supported under the Flights and Fieldwork for the Advancement of Science and Technology (FAST) funded project, ``Sounding Rocket Flight to Explore Percolating Reactive Waves,'' by the Canadian Space Agency. Computing resources used in this work were provided by Compute Canada. We thank Samuel Goroshin and Nikolas Provatas for valuable discussions, and Caroline Wagner and Francois-David Tang for performing preliminary calculations.
\end{acknowledgments}

\bibliographystyle{apsrev4-1}
\bibliography{flame_ref}

\end{document}